\documentclass[10pt,twocolumn,amsmath,amssymb,aps,floatfix,preprintnumbers,nofootinbib,nobibnotes,bibnotes]{revtex4-2}
\usepackage{graphicx}
\usepackage{dcolumn}
\usepackage{bm}
\usepackage{hyperref}
\usepackage[mathlines]{lineno}
\usepackage{epsfig}  
\usepackage{color}
\usepackage{slashed}
\usepackage{float}
\usepackage[table]{xcolor}
\usepackage{amsmath}
\usepackage{soul}
\usepackage{braket}
\usepackage{orcidlink}
\usepackage[caption=false]{subfig}

\DeclareRobustCommand{\newtheta}{\ensuremath{\Theta}}

\begin{document}

\title{Nonlocal Advantage of Quantum Coherence in Top Quarks}

\author{Saurabh Rai\,\orcidlink{0009-0004-1080-2370}}
\email{saurabhrai@iitj.ac.in}
\affiliation{Indian Institute of Technology Jodhpur, Jodhpur 342037, India}

\author{Jitendra Kumar\,\orcidlink{0000-0002-8465-433X}}
\email{jkumar@iitj.ac.in}
\affiliation{Indian Institute of Technology Jodhpur, Jodhpur 342037, India}

\begin{abstract}
There is a growing interest in investigating top-quark systems using tools from quantum information theory. A key peculiarity of the top quark is that it decays before hadronization or spin decorrelation occurs, thereby preserving its spin information. This unique property enables direct access to spin correlations, making the top quark an ideal candidate for probing fundamental quantum correlations in high-energy physics processes. A wide range of concepts from quantum information theory, such as entanglement, Bell nonlocality, quantum steering, quantum discord, and fidelity, have been investigated in this context. Several of these measures have been employed as diagnostic tools to test the Standard Model and to search for possible signatures of physics beyond. However, the \textit{nonlocal advantage of quantum coherence} (NAQC) has remained largely unexplored in this context. In this work, we present a detailed investigation of the NAQC in top quark pair production. We employ two complementary NAQC measures based on the $l_{1}$-norm and the relative entropy of coherence. We also study the effect of angular averaging on these measures and assess the sensitivity of current LHC spin-correlation measurements to NAQC. Our findings reveal rich coherence structures and highlight NAQC as potentially a novel and complementary quantum signature in high-energy physics systems.
\end{abstract}

\maketitle

\section{Introduction} 

Entanglement is one of the most striking features of quantum mechanics, encapsulating nonclassical correlations that have no counterpart in classical physics \cite{Einstein:1935rr, Bohr:1935af, Schrodinger:2008pyl, Horodecki:2009zz}. It lies at the heart of foundational tests of quantum theory as well as quantum technological applications, ranging from teleportation and superdense coding to quantum cryptography and computing \cite{Yue:2021yu}. Bell inequalities are used to explore entanglement from an operational perspective, offering measurable conditions that reveal the nonlocal nature of quantum correlations \cite{Bell:1964kc, Clauser:1969ny, Peres:1998sf, Werner:2001rf, Terhal:2000wsl}. Beyond entanglement and Bell nonlocality, a broader landscape of quantum correlations has been discovered over the years, including Einstein–Podolsky–Rosen (EPR) steering, quantum discord, contextuality, and various measures rooted in resource theories, including coherence, asymmetry, and magic states \cite{Uola:2020kps, Ollivier:2001fdq, Luo:2008ecu, Kochen:1968zz, Pavicic:2021yhu, Chitambar:2018rnj}. These correlations not only enrich our conceptual understanding of quantum theory but also provide practical advantages in tasks such as quantum communication, computation, and metrology.
 
Among the more recently explored concepts within this hierarchy is the NAQC, a distinct form of quantum correlation that establishes a connection between coherence theory and quantum nonlocality \cite{Mondal:2017prp}. It is operationally defined via a steering-type protocol, where a local measurement on one qubit can enhance the average quantum coherence in the conditional states of the other qubit beyond what is possible for any single-qubit state. Remarkably, it has been shown that any bipartite quantum state exhibiting this type of correlation must be entangled. Moreover, its presence implies a form of quantum correlation that is strictly stronger than Bell nonlocality, thereby positioning NAQC as a more stringent indicator of nonclassicality.

Several approaches have been proposed to quantify the NAQC, including basis-independent formulations and optimization over mutually unbiased bases. However, the most commonly adopted and operationally meaningful methods rely on two coherence measures: the $l_{1}$-norm of coherence and the relative entropy of coherence. Within a steering-type protocol, these measures are used to compute the average coherence of the conditional states resulting from local measurements. A bipartite state is said to exhibit NAQC when this average exceeds the maximum coherence attainable by any single-qubit state. Accordingly, these coherence-based measures provide a refined tool for probing nonclassical correlations in bipartite systems \cite{Mondal:2017prp, Hu:2018mxh}.

In the context of high-energy physics, top–antitop ($t\bar{t}$) quark pairs provide a compelling platform for investigating NAQC-like quantum correlations. These $t\bar{t}$ pairs are produced significantly at hadron colliders like the Large Hadron Collider (LHC), and due to the extremely short lifetime of the top quark ($5 \times 10^{-25}$ seconds), shorter than the hadronization timescale ($\sim 10^{-23}$ seconds), they offer a unique opportunity to probe spin correlations directly at the partonic level through their decay products. The spin degrees of freedom of the ($t\bar{t}$) system can be entangled and may encode nonclassical correlations that go even beyond Bell nonlocality. The theoretical modeling of top quark spin correlations and the manifestation of these correlations in measurable observables are discussed in detail in Ref.~\cite{D0:2011kcb, ATLAS:2012ao, ATLAS:2013gil, CMS:2013roq, ATLAS:2014aus, ATLAS:2014abv, CMS:2015cal, CMS:2016piu, ATLAS:2019zrq, Kaptanoglu:2019gtg}. Building on this, several recent studies have examined $t\bar{t}$ production through the lens of quantum information, investigating entanglement, steering, discord, and fidelity-based measures. These studies have not only deepened our understanding of the quantum structure of $t\bar{t}$ quark pair events but have also opened avenues for using such observables in probing new interactions and non-standard spin dynamics \cite{ATLAS:2023fsd, Afik:2022dgh, Afik:2022kwm, Afik:2020onf, Dong:2023xiw, Fabbrichesi:2022ovb, Han:2023fci, Maltoni:2024tul, Maltoni:2024csn, Cheng:2024btk, Cheng:2023qmz, Aguilar-Saavedra:2023hss, Aguilar-Saavedra:2024hwd, Severi:2022qjy, Aguilar-Saavedra:2024fig, White:2024nuc, Demina:2024dst, Ye:2024dqx, Han:2024ugl, CMS:2024zkc}. 

Despite these developments, a systematic study of top-antitop quark pair production is still lacking. In \cite{Afik:2022dgh}, the authors briefly discussed the existence of NAQC in the integrated $t\bar{t}$ system. In this work, we fill this gap by analyzing NAQC in leading-order $t\bar{t}$ pair production processes, viz., $q\bar{q}\to t\bar{t}$ and $gg \to t\bar{t}$. Our study extends the application of quantum information concepts to the relativistic and high-energy domain, offering a novel perspective on the nonclassical features emerging from quantum field interactions. We compute fully differential and angular averaged measures using both the $l_{1}$-norm and the relative entropy of coherence, and identify regions in the kinematic phase space where NAQC is present. Furthermore, we construct a simplified hadronic scenario by combining partonic channels with constant mixing fractions, serving as a toy model for hadron-level $t\bar{t}$ production at hadron colliders. Our findings establish NAQC as a meaningful and robust signature of quantum correlations in top-quark systems and demonstrate its potential as a complementary probe in the broader context of quantum information applied to collider physics studies.

The organisation of our paper is as follows: In Section \ref{for-tmu}, we review the formalism for modeling $t\bar{t}$ quark pair as a two-qubit system and describe the framework for investigating the NAQC in this context. Section \ref{secth} presents our main results. We start with an analysis of NAQC at the parton level using fully differential final states, followed by the study at the hadron level under a simplified toy model. Additionally, we investigate the behavior in the angular-averaged final states. Finally, Section \ref{secfo} provides a summary of our findings and outlines potential directions for future research.

\section{Formalism}
\label{for-tmu}
In this section, we outline the formalism for describing top-antitop quark pair states as a two-qubit system, introduce the framework of NAQC, and discuss its applicability to the analysis of quantum correlations in $t\bar{t}$ production.

\subsection{Top Quarks as Two-Qubit Systems}
In quantum mechanics, the state of a system is typically described by a state vector, denoted by $\ket{\psi}$ when the system is a pure state. However, in realistic scenarios, such as when the system forms part of an ensemble or when only partial information is available, this description is no longer sufficient. In such cases, the system is described using the density matrix formalism, which provides a complete statistical representation applicable to both pure and mixed states \cite{Nielsen:2012yss}. The density matrix formalism generalizes quantum states to accommodate statistical mixtures of pure states. If the system is described by an ensemble of pure states $\ket{\psi_{i}}$ occurring with probabilities $p_{i}$, the corresponding density matrix is defined as:
\begin{equation}
    \rho = \sum_{i} p_{i}\ket{\psi_{i}}\bra{\psi_{i}}
\end{equation}
where $p_{i} \geq 0$ and $\sum_{i}p_{i}=1$. However, it is not necessary to use the state-vector formalism to construct density matrices. Any generic qubit state (pure or mixed) can be written directly in the density matrix formalism as:
\begin{equation}
    \rho_{q} = \frac{\mathbb{I}_{2} + \sum_{i}\mathcal{B}_{i}\sigma^{i}}{2}
\end{equation}
where $\sigma^{i}$ are the Pauli spin matrices and $\mathbb{I}_{2}$ is the $2 \times 2$ identity matrix. The coefficients $\mathcal{B}_{i}$ encode the qubit spin polarization and must be the elements of a Bloch vector for $\rho_{q}$ to represent a physical state. The density matrix for a qubit-bipartite system, consisting of two qubit subsystems $A$ and $B$ living in Hilbert spaces $\mathcal{H}_{A}$ and $\mathcal{H}_{B}$, respectively, can be written as:
\begin{small}
\begin{equation}\label{3}
    \rho_{AB} = \frac{\mathbb{I}_{4}+\sum_{i}\mathcal{B}_{i}^{A}\sigma^{i}\otimes \mathbb{I}_{2} + \sum_{j}\mathcal{B}_{j}^{B}\mathbb{I}_{2}\otimes \sigma^{j} + \sum_{ij}\mathcal{C}_{ij}^{AB}\sigma^{i}\otimes \sigma^{j}}{4}
\end{equation}
\end{small}
where $\mathbb{I}_{4}$ is the $4 \times 4$ identity matrix, $\mathcal{B}_{i}^{A}$ and $\mathcal{B}_{j}^{B}$ represent the spin polarizations for systems $A$ and $B$, and $\mathcal{C}_{ij}^{AB}$ are the spin correlation matrix elements.

We now describe how this formalism can be applied to studying top-antitop pair production in hadron colliders \cite{Afik:2022kwm}. Consider the process
\begin{equation}
    I \to t\bar{t}
\end{equation}
where $I$ denotes an arbitrary initial state. To characterize the kinematics of the final state in the center-of-mass frame, we consider the pair’s invariant mass $m_{t\bar{t}}$, the unit vector $\hat{k}$ along one of the particle's directions (taken to be the top quark for definiteness); and the spins $s_{1}$ and $s_{2}$ associated with the top quark and antiquark, respectively. These parameters define a complete set of possible final states $\ket{m_{t\bar{t}}\hat{k}s_{1}s_{2}}$. Let \( \ket{I\lambda} \) denote an initial quantum state labeled by \( I \) with quantum numbers \( \lambda \). The amplitude for the transition from this initial state to a particular final state is then given by:
\begin{equation}
\mathcal{M}_{s_{1}s_{2}}^{I} = \braket{m_{t\bar{t}}\hat{k}s_{1}s_{2}|\hat{T}|I\lambda},
\end{equation}
where \( \hat{T} \) is the transition (scattering) matrix. From this, one constructs the \textit{R-matrix} referred to as the spin-production density matrix, which encodes the spin correlations present in the final state, and is constructed as:
\begin{equation}
R_{s_{1}s_{2},s_{1}'s_{2}'}^{I\lambda}(m_{t\bar{t}}, \hat{k}) = \mathcal{M}_{s_{1}s_{2}}^{I}(m_{t\bar{t}}, \hat{k}) \left[\mathcal{M}_{s_{1}'s_{2}'}^{I}(m_{t\bar{t}}, \hat{k})\right]^\dagger
\end{equation}

Unlike squared amplitudes used in cross sections, the \textit{R-matrix} retains spin labels in both the initial and final states. In experimental observations, quantum numbers such as color and spin are typically not resolved. Therefore, we define a spin-averaged \textit{production density matrix} as:
\begin{equation}
R_{s_{1}s_{2},s_{1}'s_{2}'}^I(m_{t\bar{t}}, \hat{k}) = \frac{1}{N_\lambda} \sum_{\lambda} R_{s_{1}s_{2},s_{1}'s_{2}'}^{I\lambda}(m_{t\bar{t}}, \hat{k}),
\end{equation}
which can be normalized to obtain a proper quantum density matrix with unit trace:
\begin{equation}
\rho^I = \frac{R^I}{\text{Tr}(R^I)}
\end{equation}

This matrix \( \rho^I \) captures the spin state of the top-antitop quark pair for fixed \( m_{t\bar{t}} \) and \( \hat{k} \). Yet, one might also be interested in the \textit{mixed spin density matrix} of the top quark system, averaged over all possible directions \( \hat{k} \) for the fixed invariant mass. This leads to the following construction:
\begin{small}
\begin{equation}
\bar{\rho}^I(m_{t\bar{t}}) = \frac{1}{Z} \sum_{s_{1}s_{2},s_{1}'s_{2}'} \int d\Omega\, R^I(m_{t\bar{t}}, \hat{k}) \frac{\ket{m_{t\bar{t}}\hat{k}s_{1}s_{2}}\bra{m_{t\bar{t}}\hat{k}s_{1}'s_{2}'}}{\braket{m_{t\bar{t}}\hat{k}|m_{t\bar{t}}\hat{k}}}
\end{equation}
\end{small}
Here, $R^{I} (m_{t\bar{t}}, \hat{k})$ is the shorthand for $R_{s_{1}s_{2}, s_{1}'s_{2}'}^{I} (m_{t\bar{t}}, \hat{k})$.
This integral spans all solid angles \( \Omega \), and the denominator normalizes each contribution. The prefactor \( Z \) ensures total normalization. Now, it is known that for any two-qubit system, such as the top-antitop quark pair, the R-matrix can be decomposed as:
\begin{small}
\begin{equation}
R^I = \tilde{A}^I \mathbb{I}_4 + \sum_i \left( \tilde{B}_i^{I+} \sigma_i \otimes \mathbb{I}_2 + \tilde{B}_i^{I-} \mathbb{I}_2 \otimes \sigma_i \right) + \sum_{ij} \tilde{C}_{ij}^I \sigma_i \otimes \sigma_j.
\end{equation}
\end{small}
Here, \( \mathbb{I}_n \) is the identity matrix in \( n \)-dimensions and \( \sigma_i \) are the Pauli matrices. The kinematic details are embedded in the coefficients \( \tilde{A}^I, \tilde{B}_i^{I\pm}, \tilde{C}_{ij}^I \). Upon normalization via the trace operation, only the scalar term \( \tilde{A}^I \) survives, yielding:
\begin{equation}
\rho^I = \frac{R^I}{4\tilde{A}^I}.
\end{equation}

Consequently, the normalized matrix \( \rho^I \) has the decomposition:
\begin{equation}
B_i^{I\pm} = \frac{\tilde{B}_i^{I\pm}}{\tilde{A}^I}, \quad C_{ij}^I = \frac{\tilde{C}_{ij}^I}{\tilde{A}^I}.
\end{equation}
These coefficients can then be directly matched to the ones in Eq.~\ref{3}.
To evaluate them explicitly, a specific coordinate system must be chosen, and a standard choice is the \textit{helicity basis}, where directions transverse to the top-quark momentum \( \hat{k} \) are defined as:

\begin{equation}
\hat{r} = \frac{\hat{p} - \cos\Theta\, \hat{k}}{\sin\Theta}, \quad \hat{n} = \hat{r} \times \hat{k},
\end{equation}

with \( \hat{p} \) being the direction of the initial beam and \( \Theta \) the scattering angle (\( \cos\Theta = \hat{k} \cdot \hat{p} \)) \cite{Mahlon:1995zn, Brandenburg:1996df}.

One advantage of the helicity frame is that Lorentz boosts along the direction \( \hat{k} \) (defined by the top-quark momentum in the center-of-mass frame) leave the spin quantization axis invariant. This property makes spin correlations readily interpretable in both the rest frame and the laboratory frame. Another interesting choice of spin quantization axis is the \textit{off-diagonal basis} \cite{Mahlon:1997uc}, which maximizes the spin asymmetry, particularly for the $q\bar{q} \to t\bar{t}$ channel and thereby enhances the observable angular correlations between the decay products. In this basis, the scattering amplitudes for like-spin configurations of the top–antitop pair vanish identically at leading order. In the limit \( \beta \to 1 \), where the top quark approaches its relativistic regime, this basis converges to the helicity basis. Expressions for the polarization and correlation coefficients in both bases can be found in Refs.~\cite{Brandenburg:1996df, Afik:2022kwm} and are also included here for completeness.

\textit{Production spin density matrix elements:} As a general feature, top quark states are known to be Bell-diagonal \cite{Afik:2022kwm, CMS:2024zkc, CMS:2019nrx}. We begin by presenting the production spin density matrix elements in the helicity basis. For the \( q\bar{q} \) process, the matrix elements are listed in Eq.~\ref{eq:14}, whereas for the \( gg \to t \bar{t} \) process, the matrix elements are given in Eq.~\ref{eq:15}.

\begin{equation}
\begin{aligned}
\tilde{A}^{q\bar{q}} &= F_{q\bar{q}}\left(2 - \beta^2\sin^2\Theta\right) \\
\tilde{C}^{q\bar{q}}_{rr} &= F_{q\bar{q}}\left(2 - \beta^2\right)\sin^2\Theta \\
\tilde{C}^{q\bar{q}}_{nn} &= -F_{q\bar{q}}\beta^2\sin^2\Theta \\
\tilde{C}^{q\bar{q}}_{kk} &= F_{q\bar{q}}\left[2 - (2 - \beta^2)\sin^2\Theta\right] \\
\tilde{C}^{q\bar{q}}_{rk} &= \tilde{C}^{q\bar{q}}_{kr} = F_{q\bar{q}}\sqrt{1-\beta^2}\sin 2\Theta \\
F_{q\bar{q}} &= \frac{1}{18}
\end{aligned}
\label{eq:14}
\end{equation}

\begin{equation}
\begin{aligned}
\tilde{A}^{gg} &= F_{gg}\left[1 + 2\beta^2\sin^2\Theta - \beta^4\left(1 + \sin^4\Theta\right)\right] \\
\tilde{C}^{gg}_{rr} &= -F_{gg}\left[1 - \beta^2(2-\beta^2)(1+\sin^4\Theta)\right] \\
\tilde{C}^{gg}_{nn} &= -F_{gg}\left[1 - 2\beta^2 + \beta^4(1+\sin^4\Theta)\right] \\
\tilde{C}^{gg}_{kk} &= -F_{gg}\left[1 - \beta^2\frac{\sin^2 2\Theta}{2} - \beta^4(1+\sin^4\Theta)\right] \\
\tilde{C}^{gg}_{rk} &= \tilde{C}^{gg}_{kr} = F_{gg}\sqrt{1-\beta^2}\beta^2\sin 2\Theta\sin^2\Theta \\
F_{gg} &= \frac{7 + 9\beta^2\cos^2\Theta}{192(1-\beta^2\cos^2\Theta)^2}
\end{aligned}
\label{eq:15}
\end{equation}
where,
\begin{equation}
    \beta = \sqrt{1-\frac{4m_{t}^{2}}{m_{t\bar{t}}^{2}}}
\end{equation}
With appropriate rotations, the spin correlation matrix elements in the helicity basis can be converted into those in the off-diagonal basis. In this basis, for $q\bar{q} \to t\bar{t}$ process, we have the diagonal elements as:
\begin{equation}
    \begin{aligned}
        C^{q\bar{q}}_{1} &= 1 \\
        C^{q\bar{q}}_{2} &= \frac{-\beta^{2}\sin^{2}\theta}{2-\beta^{2}\sin^{2}\theta} \\
        C^{q\bar{q}}_{3} &= -C^{q\bar{q}}_{2}
    \end{aligned}
\end{equation}
while for $gg \to t\bar{t}$ process:
\begin{widetext}
\begin{equation}
\begin{aligned}
C^{gg}_{1} &= 
\dfrac{-1+\beta^2(1+\sin^2\newtheta) 
+ \sqrt{\beta^4(1-2\sin^2\newtheta+5\sin^4\newtheta) - 2\beta^6(1-\sin^2\newtheta+3\sin^4\newtheta+\sin^6\newtheta) + \beta^8(1+\sin^4\newtheta)^2}}
{1+2\beta^2\sin^2\newtheta-\beta^4(1+\sin^4\newtheta)}\\
C^{gg}_{2} &= -\frac{1 - 2\beta^2 + \beta^4(1+\sin^4\newtheta)}{1 + 2\beta^2\sin^2\newtheta - \beta^4\left(1 + \sin^4\newtheta\right)}\\
C^{gg}_{3} &= \dfrac{-1+\beta^2(1+\sin^2\newtheta) 
- \sqrt{\beta^4(1-2\sin^2\newtheta+5\sin^4\newtheta) - 2\beta^6(1-\sin^2\newtheta+3\sin^4\newtheta+\sin^6\newtheta) + \beta^8(1+\sin^4\newtheta)^2}}
{1+2\beta^2\sin^2\newtheta-\beta^4(1+\sin^4\newtheta)}
\end{aligned}
\end{equation}
\end{widetext}
\subsection{Nonlocal Advantage of Quantum Coherence}
Quantum coherence, a fundamental feature of quantum mechanics, describes the ability of a system to exhibit interference. In bipartite systems, coherence can also manifest nonlocally. Specifically, when a measurement on one part of the system (e.g., Alice's qubit) results in conditional states on the other part (e.g., Bob's qubit) that display coherence beyond what any single qubit can achieve alone, the system is said to exhibit NAQC \cite{Mondal:2017prp}. This phenomenon highlights a distinct form of quantum correlation, wherein the coherence observed in Bob’s subsystem cannot be ascribed to local properties alone. Instead, it arises from the global quantum nature of the composite system. Notably, any quantum state that exhibits NAQC is necessarily entangled and also demonstrates Bell nonlocality.

To detect NAQC, a well-defined protocol is enacted where Alice measures her qubit in one of the Pauli bases \( \{\sigma_1, \sigma_2, \sigma_3\} \), and communicates her outcome to Bob. Bob then measures coherence of his conditional state in a basis different from the one used by Alice, allowing the identification of coherence that surpasses local limitations and reveals the nonlocal quantum correlations inherent in the system. We discuss two measures of NAQC based on the $l_{1}$-norm of coherence and the relative entropy of coherence.

\textit{The \( l_1 \)-Norm of Coherence}: This measure is defined as the sum of off-diagonal absolute values of the density matrix in a given basis and evaluated as:
\begin{equation}
C^{\text{na}}_{l_1}(\rho) = \frac{1}{2} \sum_{i \neq j} \sum_{a = 0}^1 p(a|\sigma_i) \, C^{\sigma_j}_{l_1}(\rho_{B|\sigma_i^a}),
\end{equation}
where \( p(a|\sigma_i) \) is the probability of Alice obtaining outcome \( a \) when measuring observable \( \sigma_i \), and \( \rho_{B|\sigma_i^a} \) is the corresponding conditional state of Bob. A state exhibits NAQC~\cite{Mondal:2017prp} if:
\begin{equation}
C^{\text{na}}_{l_1}(\rho) > C_{l_1}^{\text{crit}} = \sqrt{6} \approx 2.449.
\end{equation}
\\
Consider the Bell-diagonal states \( \rho_{\text{Bell}} \) expressed as: 
\[
\rho_{\text{Bell}} = \frac{1}{4} \left( I \otimes I + \sum_{i=1}^{3} v_i \, \sigma_i \otimes \sigma_i \right)
\]
where $\{v_{i}\}$ is the correlation vector.

The expression of $C_{l_{1}}^{na}$ in this case simplifies to:
\begin{equation}
C^{\text{na}}_{l_1}(\rho_{\text{Bell}}) = |v_1| + |v_2| + |v_3|.
\end{equation}

\textit{Relative Entropy of Coherence}: An alternative quantifier of coherence is the relative entropy, which quantifies the distinguishability between a quantum state and its fully decohered counterpart. We construct this measure as follows:
\begin{equation}
C^{\text{na}}_{\text{re}}(\rho) = \frac{1}{2} \sum_{i \neq j} \sum_{a = 0}^1 p(a|\sigma_i) \, C^{\sigma_j}_{\text{re}}(\rho_{B|\sigma_i^a}),
\end{equation}
The coherence of a quantum state  $\rho$ in the basis \( \sigma_j \) is quantified by the relative entropy of coherence, defined as:
\begin{equation}
C^{\sigma_j}_{\text{re}}(\rho) = S(\rho^{\text{diag}}_{\sigma_j}) - S(\rho),
\end{equation}
where \( S(\cdot) \) denotes the von Neumann entropy and \( \rho^{\text{diag}}_{\sigma_j} \) is the state obtained by retaining only the diagonal elements of $\rho$ in the eigen-basis of \( \sigma_j \). The corresponding threshold for demonstrating NAQC using this measure is given by ~\cite{Mondal:2017prp}:
\begin{equation}
C^{\text{na}}_{\text{re}}(\rho) > C_{\text{re}}^{\text{crit}} \approx 2.232.
\end{equation}
For Bell-diagonal states \( \rho_{\text{Bell}} \), the expression simplifies to a closed-form that allows direct evaluation of the NAQC condition ~\cite{Hu:2018mxh}:
\begin{equation}
C^{\text{na}}_{\text{re}}(\rho_{\text{Bell}}) = 3 - \sum_{i=1}^{3} H\left( \frac{1 + v_i}{2} \right),
\end{equation}
where \( H(p) = -p \log_2 p - (1 - p) \log_2 (1 - p) \) is the binary Shannon entropy. This closed form applies only when the spin-correlation matrix is diagonal, which is the case when the off-diagonal basis is used for measuring the matrix elements. In experimental settings at the LHC, however, the helicity basis is often used instead, introducing off-diagonal correlations; in such cases, one must treat the full, generic two-qubit state. To the best of the authors' knowledge, no closed-form expression for the NAQC quantified via the $l_{1}$ norm in the helicity basis, that is, one applicable to an arbitrary two-qubit state, has been reported in the literature. For a generic correlation matrix, one can only conclude when NAQC is absent; in other regions of the parameter space, the existing criteria do not provide a definite conclusion. In contrast, for the relative-entropy formulation of NAQC, a closed-form expression is available for states with vanishing polarization. For this class of states, one has:
\begin{equation}
    C_{re}^{na} = \frac{1}{2} \sum_{i \neq j} H\!\left(\frac{1 + C_{ij}}{2}\right)
- \sum_{i} H\!\left(\frac{1 + \sqrt{\sum_{j} C_{ij}^2}}{2}\right),
\end{equation}
In the theoretical analyses that follow, we focus on the off-diagonal basis for studying NAQC in top-quark systems. In addition to the obvious limitation of achieving concrete analytical expressions in the helicity basis, this choice is primarily motivated by the fact that nonlocal features and coherence properties are more pronounced and intuitive to understand in the off-diagonal basis \cite{Hu:2018mxh}. We revisit the helicity basis in Section~\ref{sec:4}, where we examine its relevance in light of the recent measurement reported by the CMS Collaboration in~\cite{CMS:2024zkc}.


\section{NAQC in top quarks}
\label{secth}
\subsection{Fully differential final states}
\subsubsection{NAQC at parton level}
We begin by investigating the presence of NAQC in top–antitop quark pair production arising from partonic initial states, with full differential dependence on the kinematic variables. Specifically, the production density matrices are considered as functions of both the scattering angle $\theta$ and the invariant mass of the top-quark pair.

\begin{figure}[!htb]
    \centering
    \includegraphics[width=\linewidth]{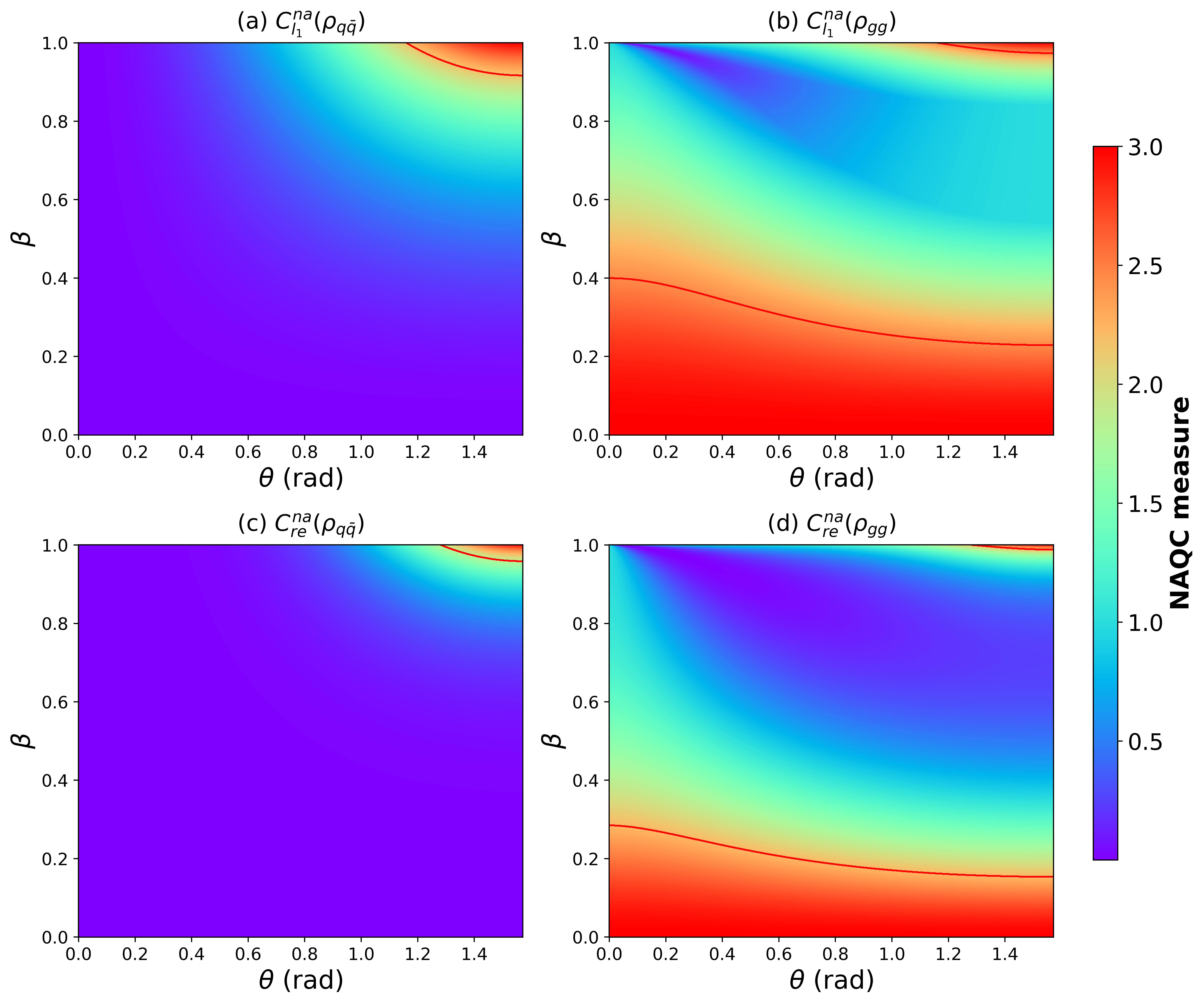}
    \caption{NAQC measure in the \( q\bar{q} \)-initiated process (left panel) and the \( gg \)-initiated process (right panel) for top-antitop quark pair production. The top panel shows the measure based on the \( l_1 \)-norm of coherence (\( C^{\text{na}}_{l_1} \)), while the bottom panel shows the measure based on the relative entropy (\( C^{\text{na}}_{\text{re}} \)).}
    \label{fig:naqc-hierarchy}
\end{figure}

Fig.~\ref{fig:naqc-hierarchy} illustrates the NAQC landscape in top-quark pair production for the $q\bar{q}$ and $gg$ initial-state channels via the ${l_1}$-norm of coherence and the relative entropy of coherence. The red contour identifies the regions in kinematic space where the criteria for NAQC are satisfied, whereas the black contour marks the subset of these regions that also exhibit Bell nonlocality. 

The top panel illustrates the NAQC, quantified via the \( C^{\mathrm{na}}_{l_1} \) measure, for the processes \( q\bar{q} \to t\bar{t} \) and \( gg \to t\bar{t} \), respectively. In the top-left plot, corresponding to the quark–antiquark annihilation channel, NAQC is largely absent across the kinematic phase space, except within a narrow region characterized by large scattering angles (\( \theta \to \pi/2 \)) and high boosts (\( \beta \to 1 \)). In contrast, the top-right plot, associated with the gluon–gluon fusion channel, exhibits a more pronounced presence of NAQC over a broader range of both low and high values of \( \beta \) and \( \theta \).

It is important to note that while NAQC is always accompanied by Bell nonlocality, the converse does not hold. This is consistent with the established hierarchy of quantum correlations, where NAQC constitutes a stronger form of quantum correlation than Bell nonlocality~\cite{Mondal:2017prp}.\\
Furthermore, although previous studies have shown that the Bell inequality may be violated across a broad region of phase space in the \( q\bar{q} \to t\bar{t} \) process~\cite{Fabbrichesi:2021npl}, our results indicate that NAQC emerges only within a restricted subset of this phase space.

Similarly, the bottom panel illustrate the \( C^{\mathrm{na}}_{\mathrm{re}} \) measure. The overall behavior mirrors that observed with the \( l_1 \)-norm measure; however, the regions where NAQC is achieved are noticeably reduced. This outcome aligns with expectations, as the \( l_1 \)-norm of coherence typically identifies a broader subset of states exhibiting NAQC compared to the relative entropy measure~\cite{Hu:2018mxh}.
\subsubsection{NAQC at hadronic production level} In high-energy collider experiments such as the LHC, fundamental interactions take place between partons confined within composite hadrons. Consequently, any observable related to final-state quantum correlations, such as entanglement, coherence, or NAQC, must be constructed at the hadronic level, properly accounting for the composite and probabilistic nature of the initial state. In this context, partonic subprocesses such as \( q\bar{q} \to t\bar{t} \) and \( gg \to t\bar{t} \) are embedded within the hadronic reaction, and their respective contributions are weighted by \textit{parton distribution functions} (PDFs), which encode the probability of finding a parton carrying a specific momentum fraction at a given scale.

The \textit{hadronic density matrix} for top-quark pair production can be written as a weighted sum of partonic density matrices:
\[
\rho(M_{t\bar{t}}, \hat{k}, \sqrt{s}) = \sum_{I=q\bar{q}, gg} w_I(M_{t\bar{t}}, \hat{k}, \sqrt{s}) \, \rho^I(M_{t\bar{t}}, \hat{k}),
\]
where each \( \rho^I \) corresponds to the density matrix associated with a specific partonic initial state \( I \), and the weights \( w_I \) are determined by a combination of PDF-derived \textit{luminosity functions} \( L_I \) and kinematic matrix elements \( \tilde{A}^I \). These weights are given by
\[
w_I(M_{t\bar{t}}, \hat{k}, \sqrt{s}) = \frac{L_I(M_{t\bar{t}}, \sqrt{s}) \, \tilde{A}^I(M_{t\bar{t}}, \hat{k})}{\sum_J L_J(M_{t\bar{t}}, \sqrt{s}) \, \tilde{A}^J(M_{t\bar{t}}, \hat{k})},
\]
where \( L_I \) encodes the \textit{partonic luminosity}, integrated over the appropriate PDF combinations, and \( \tilde{A}^I \) contains the angular and helicity information for the production amplitude squared.

In principle, this expression provides the \textit{most accurate description} of the hadronic-level quantum state in terms of partonic inputs and hadron structure. However, to isolate and understand the dependence of quantum coherence on initial-state composition, we adopt a \textit{simplified toy model} where the hadronic density matrix is approximated as a convex combination of the two dominant channels:
\[
\rho_{\text{had}} = f \, \rho_{q\bar{q}} + (1 - f) \, \rho_{gg},
\]
where \( f \) is treated as a constant mixing parameter, \textit{independent of} \( \theta \), \( \beta \), or \( m_{t\bar{t}} \). This simplification effectively ignores the dynamical role of PDFs and parton luminosities, allowing us to explore, in a controlled fashion, how quantum observables such as NAQC vary as a function of subprocess composition.

\begin{figure}[h!]
    \centering
    \includegraphics[width=0.98\linewidth]{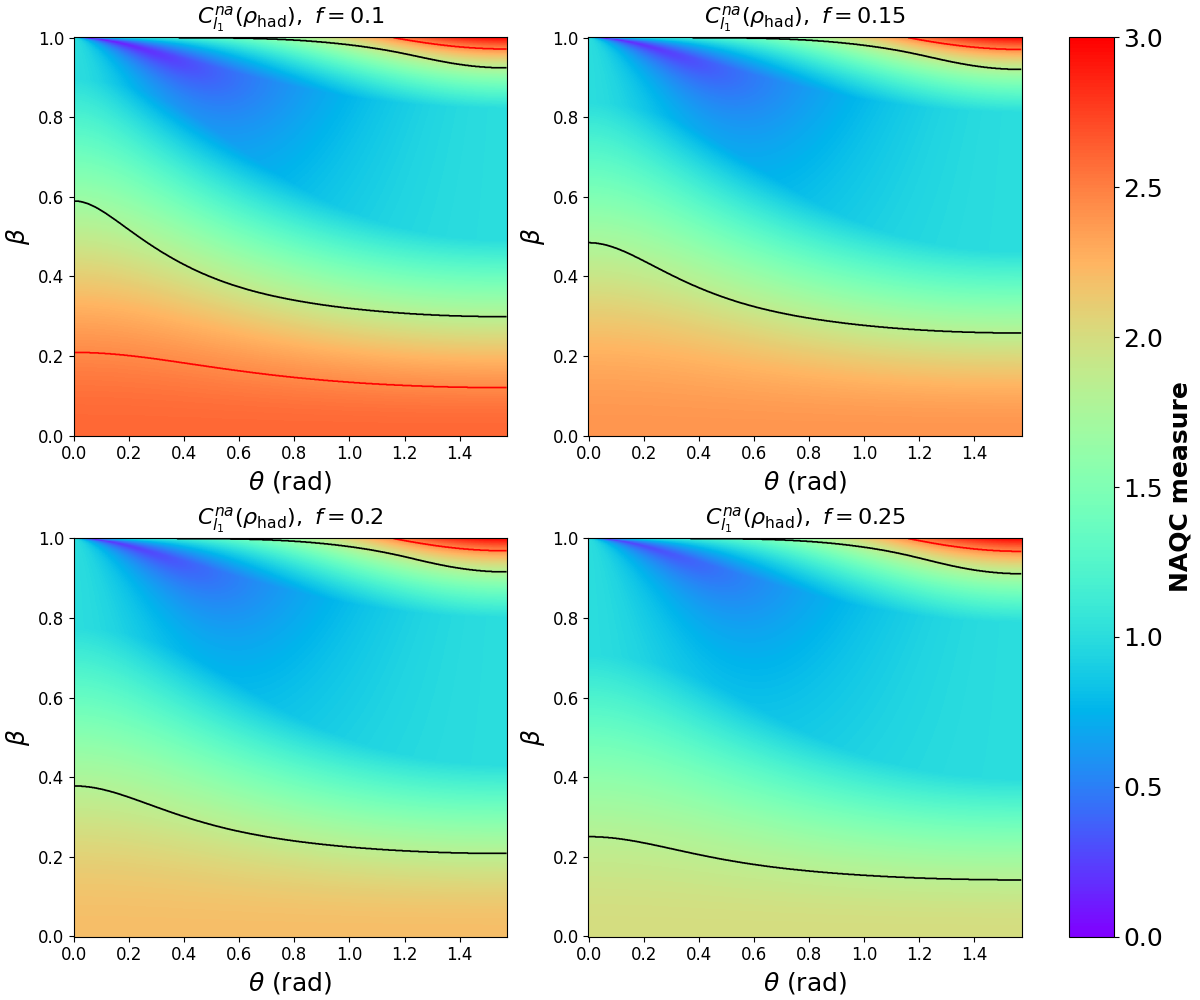}
    \caption{Investigation of NAQC in top-antitop quark pair production using a toy hadronic process with various mixing parameters $f$. The black contours mark the regions showing Bell-nonlocality.}
    \label{fig:hadron1}
\end{figure}

Fig.~\ref{fig:hadron1} illustrates the behavior of \( C_{l_1}^{na}(\rho_{\text{had}}) \) for different values of the mixing parameter \( f \) ranging from 0.1 to 0.25. We observe that as the parameter \( f \) increases, the region exhibiting NAQC, indicated by the red contours, shrinks steadily. At lower values of \( f \), particularly \( f = 0.1 \), quantum coherence is predominantly inherited from the gluon-fusion process, giving rise to NAQC across a wide portion of the phase space \((\theta, \beta)\). For higher values of \( f \), this coherence is progressively diluted by the less-structured \( q\bar{q} \) contribution, leading to a suppression of both NAQC and Bell nonlocality across the phase space \((\theta, \beta)\).

It is important, however, to acknowledge the interpretational limitations of this toy model. The use of a fixed fraction \( f \) across all points in phase space does not accurately capture the physical complexity of hadronic collisions, where the relative contribution of each partonic subprocess varies with \( \hat{k} \), rapidity, and transverse momentum, as determined by the parton distribution functions (PDFs). Moreover, the \textit{absence of NAQC} in certain regions should not be misinterpreted as evidence of classicality or a lack of quantum correlations. Rather, it signifies that the underlying coherence, if present, is insufficient to surpass the NAQC threshold under the specific quantifier employed (e.g., \( C_{l_1}^{\mathrm{na}} \) or \( C_{\mathrm{re}}^{\mathrm{na}} \)).

In summary, while the simplified \( f \)-dependent model offers qualitative insights into how quantum coherence and entanglement respond to varying initial-state compositions, a fully realistic analysis would require implementing the expressions for \( w_I \) in terms of PDFs and partonic amplitudes to construct the dynamically weighted hadronic density matrix. This would enable a more accurate mapping of quantum information measures across collider kinematics and pave the way for experimentally viable predictions.

At the LHC, the relative contributions of \( q\bar{q} \to t\bar{t} \) and \( gg \to t\bar{t} \) are well understood within the Standard Model, with the \( gg \) fusion initiated process accounting for approximately $90\%$ of the total cross-section at the centre-of-mass energy of around 13 TeV \cite{Czakon:2013tha}, corresponding to an effective mixing fraction ($f$) of 0.1 in our toy model. However, in the presence of new physics, such as the exchange of heavy resonances, contact interactions, or the modification of top-quark couplings, this balance could be altered. Since the structure of quantum correlations, including NAQC, is sensitive to the production mechanism, such changes could leave discernible imprints on coherence-based observables. Thus, NAQC may offer a novel and complementary diagnostic tool for identifying deviations from Standard Model dynamics in top pair production.

\subsection{Angular-averaged final states}
\label{obs-bk}
As we have seen, the spin density matrix \(\rho^I\) describes the quantum state of the top-antitop quark system for fixed invariant mass \(m_{t\bar{t}}\) and a given production direction \(\hat{k}\). However, in many realistic scenarios, particularly when angular resolution is limited, it is meaningful to consider the mixed spin state obtained by averaging over all possible production angles. This leads to the construction of the angular-averaged spin density matrix \(\bar{\rho}^I(m_{t\bar{t}})\), defined in Eq.~(9), which integrates the production density matrix \(R^I\) over solid angle and includes spin sums weighted by the overlap of momentum eigenstates. This averaged object provides a natural representation of the final state in the absence of directional information, such as when observables are inclusive over angular variables.

\begin{figure}[!hbt]
    \centering
    \includegraphics[width=0.90\linewidth]{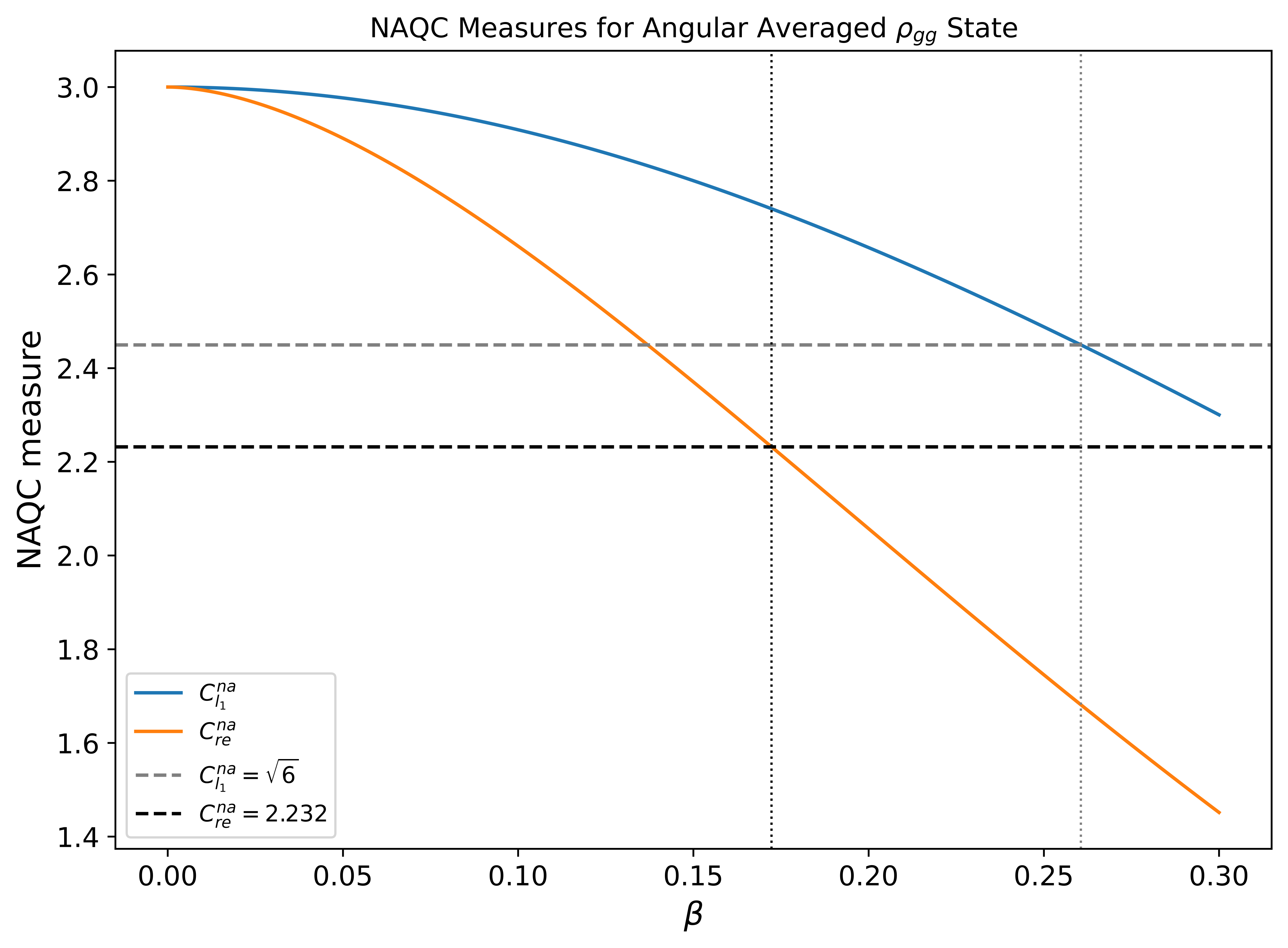}
    \caption{NAQC measures for angular-averaged $\rho_{gg}$ state. The horizontal gray and black dashed lines indicate the threshold values of NAQC based on the \( l_{1} \)-norm and relative entropy measures, respectively. The vertical dotted lines mark the corresponding bounds on \( \beta \) for each measure.}
    \label{fig:Integrated-NAQC}
\end{figure}

The behavior of the NAQC measures evaluated for the angular-averaged final state exhibits pronounced differences between the two dominant partonic production mechanisms, as illustrated in Fig.~\ref{fig:Integrated-NAQC}. For the \( q\bar{q} \to t\bar{t} \) process, both the \( C_{l_1}^{\text{na}} \) and \( C_{\mathrm{re}}^{\text{na}} \) measures remain consistently below unity across the entire range of top-quark velocities \( \beta \), with no instance of the NAQC threshold being surpassed. This indicates that, even after integrating over angular variables, the coherence generated in the final state from quark–antiquark annihilation channel remains limited and does not reach the level necessary to exhibit the operational signature of NAQC. On the other hand, the \(gg \to t\bar{t}\) process exhibits significant NAQC, with both coherence measures starting at maximal values and decreasing with \(\beta\) while still exceeding their classical bounds over a broad region.

The \( l_{1} \)-norm based measure \( C_{l_{1}}^{na} \) indicates nonlocal advantage for \( \beta \in [0, 0.26] \), while the entropic measure does so for \( \beta \in [0, 0.17] \). As noted earlier, the entropic criterion for NAQC is more stringent, and fewer final states satisfy it based on relative entropy. For angular-averaged states, the correlation matrix becomes diagonal in both the off-diagonal and helicity bases, which allows the determination of the $\beta$ bounds corresponding to NAQC in each case. Interestingly, the same $\beta$ bounds are obtained in both bases, indicating that the observation of NAQC is basis-independent under angular averaging.

\section{Probing NAQC using LHC Data}\label{sec:4}
\begin{figure}[h]
    \centering
    \includegraphics[width=0.98\linewidth]{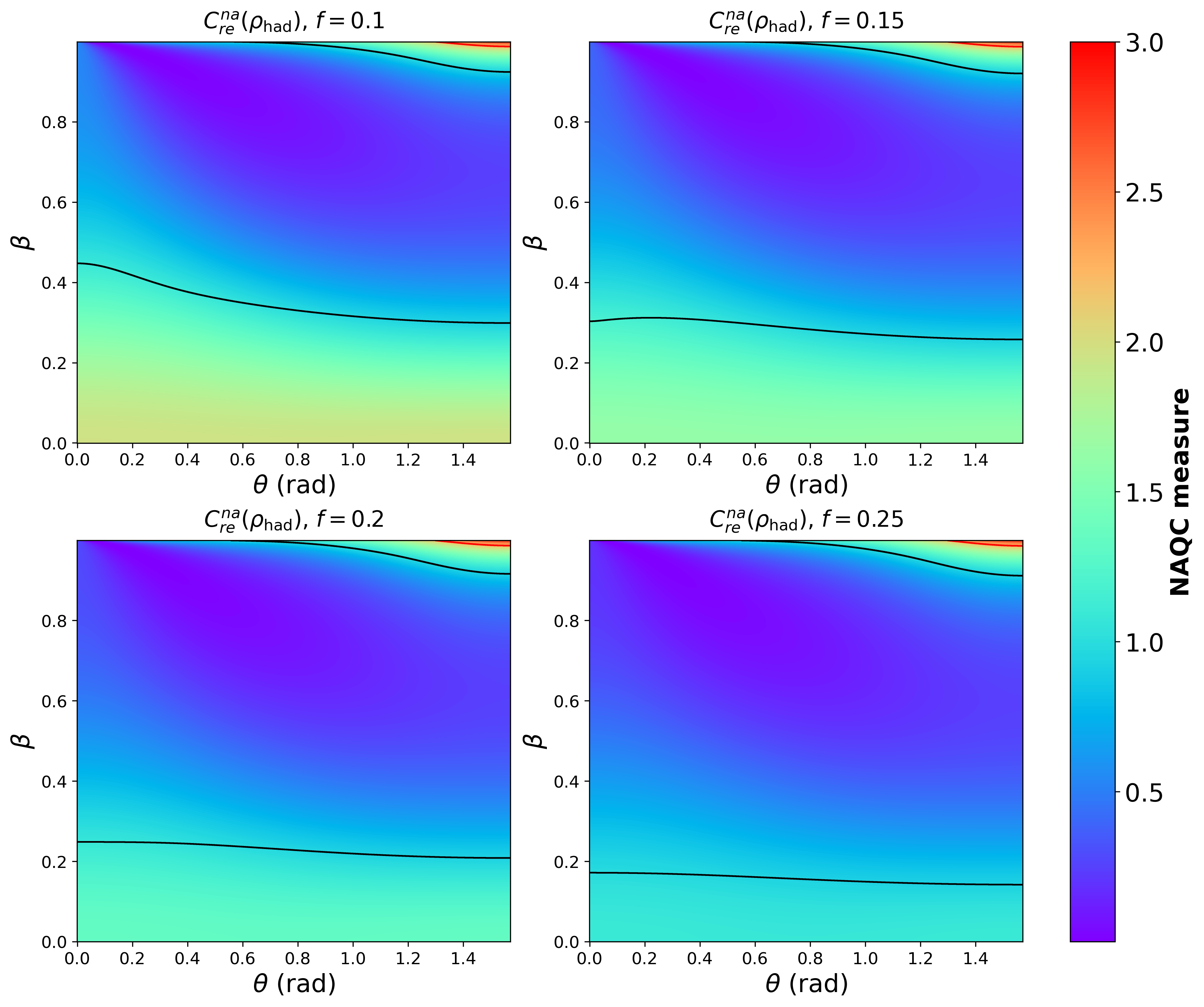}
    \caption{Investigation of NAQC based on relative entropy in top-antitop quark pair production using a toy hadronic process with various mixing parameters $f$ (helicity basis). The black contours mark the regions showing Bell-nonlocality.}
    \label{figure4}
\end{figure}

As discussed earlier, the CMS Collaboration reports the spin–correlation matrix elements for top–pair production in the helicity basis. To place our relative-entropy of coherence analysis in direct correspondence with these measurements, we also present a plot analogous to Fig.~\ref{fig:hadron1}, as Fig.~\ref{figure4}. As the plot shows, except in the very high-energy regime, the measured correlation matrix in the helicity basis is not expected to yield a detectable NAQC signal.

To further validate this, we perform an experimentally constrained, data-driven pseudo-experiment based on the spin–correlation matrix elements reported by the CMS Collaboration for the relevant phase–space regions (e.g., $m_{t\bar t} > 800~\text{GeV}$, $|\cos\theta| < 0.4$) in Ref.~\cite{CMS:2024zkc}. We generate ensembles of pseudo-experiments by sampling the correlation coefficients $C_{ij}$ from independent Gaussian distributions centered on the CMS central values with widths given by the corresponding uncertainties. For each pseudo-experiment, we compute the quantity $C^{na}_{re}$ and compare it with the critical threshold $C^{na}_{re,\text{crit}} \simeq 2.232$. A total of $2\times 10^5$ independent pseudo–experiments were performed. The distribution of $C^{na}_{re}$ was then summarized by its mean, standard deviation, and the empirical fraction of samples exceeding the critical threshold, with binomial confidence intervals estimated via the Clopper–Pearson method. 

In addition, we applied a one–sided $t$–test to evaluate the null hypothesis $H_0:\ \langle C^{na}_{re}\rangle \le C^{na}_{re,\text{crit}}$ against the alternative $H_1:\ \langle C^{na}_{re}\rangle > C^{na}_{re,\text{crit}}$. We obtain \[\langle C^{na}_{re}\rangle = 2.12 \pm 0.11,\] with only about $14.8\%$ of pseudo–experiments exceeding the critical threshold. The one–sided test yields $p \simeq 1$, indicating no statistical support for $C^{na}_{re} > C^{na}_{re,\text{crit}}$. We therefore conclude that, based on the current CMS measurements of the spin–correlation matrix in the helicity basis~\cite{CMS:2024zkc} and within present experimental uncertainties, there is no statistically significant evidence for NAQC in $t\bar{t}$ production at 13~TeV. We also checked all the other regions of phase-space and were unable to reject the above null hypothesis. Because the $l_{1}$-norm criterion is strictly weaker than the relative-entropy criterion, the absence of NAQC according to $C^{na}_{\mathrm{re}}$ implies that $C^{na}_{l_{1}}$ cannot signal NAQC either. However, since quantum-correlation measures are strongly basis dependent, and our analysis shows that the off-diagonal basis is substantially more sensitive to NAQC, measurements performed in this basis may reveal signatures of NAQC even at current LHC energies.


\section{Conclusion}
\label{secfo}
\label{conc-gen}
In this work, we have presented the first comprehensive analysis of the NAQC in the top-antitop quark system. As discussed, NAQC lies at the highest in the hierarchy of quantum correlations, with the relation \( \mathcal{N} \subset \mathcal{B} \). We establish this hierarchy in the $t\bar{t}$ system using two different measures. Furthermore, we find that NAQC is pronounced in a broad region of the phase space (\( \theta, \beta \)) in the gluon-gluon fusion initiated process alone. This distinctive behavior suggests that the NAQC may serve as a potential probe for testing the Standard Model and exploring possible signatures of physics beyond it. For angular-averaged final states, we found that although NAQC is diluted, it is independent of the choice of basis. Based on data-driven pseudo-experiment studies using the published CMS helicity-basis results, we find no indication of detectable NAQC within the reported uncertainties. However, our analysis demonstrates that the off-diagonal basis offers better sensitivity, indicating that measurements performed in this basis could reveal signs of  NAQC even at present LHC energies. Future studies may also conduct a more detailed analysis of NAQC in realistic collider environments, incorporating detector effects and backgrounds, as well as investigating its sensitivity to scenarios for testing SM and identifying signatures of New Physics.

\section{Data Availability}
The data that support the findings of this article are openly available in Ref.~\cite{datanaqc}

\vspace{0.1in}
\section{Acknowledgement}
The authors acknowledge financial support from the DST-SERB, India, under Grant No. EEQ/2023/000959, and from the I.I.T. Jodhpur, India, under Project No. I/RIG/JTK/20240067.

\end{document}